%% file: Designing_for_Understanding_How_Interface-Level_Consent_Designs_Shape_A.tex
\documentclass[sigconf]{acmart}
\usepackage{threeparttable}
\usepackage{multirow}
\usepackage{multicol}
\usepackage{placeins}
\usepackage{float}
\AtBeginDocument{%
  }

\copyrightyear{2026}
\acmYear{2026}
\setcopyright{cc}
\setcctype{by}
\acmConference[CHI EA '26]{Extended Abstracts of the 2026 CHI Conference on Human Factors in Computing Systems}{April 13--17, 2026}{Barcelona, Spain}
\acmBooktitle{Extended Abstracts of the 2026 CHI Conference on Human Factors in Computing Systems (CHI EA '26), April 13--17, 2026, Barcelona, Spain}
\acmDOI{10.1145/3772363.3798884}
\acmISBN{979-8-4007-2281-3/2026/04}

\usepackage{enumitem}

\begin{document}
\input{questionsformatting}

\title[Designing for Understanding]{Designing for Understanding: How Interface-Level Consent Designs Shape Attention and Understanding in Privacy Disclosures}


\author{Wei Xiao}
\email{weixiao8@illniois.edu}
\affiliation{
    \institution{University of Illinois Urbana-Champaign}
    \department{Institute of Communications Research}
    \city{Champaign}
    \state{Illinois}
    \country{USA}
}

\author{Mengke Wu}
\email{mengkew2@illinois.edu}
\affiliation{
    \institution{University of Illinois Urbana-Champaign}
    \department{School of Information Sciences}
    \city{Champaign}
    \state{Illinois}
    \country{USA}
}

\author{Yeeun Jo}
\email{yeeunjo2@illinois.edu}
\affiliation{
    \institution{University of Illinois Urbana-Champaign}
    \department{Siebel School of Computing and Data Science}
    \city{Champaign}
    \state{Illinois}
    \country{USA}
}


\begin{abstract}
Privacy policies are intended to support informed consent, yet users rarely read them fully. This study examines how common privacy policy interface structures influence attention allocation, reading behavior, and perceived experience. Using eye-tracking and post-task surveys, we compared three interface designs: continuous scrolling text, collapsible sections, and collapsible sections with brief previews. Results show that interface structure systematically shaped how users allocated attention and navigated policy content, but did not uniformly improve comprehension. Guided layouts supported more efficient and coherent reading patterns, whereas more interactive designs elicited higher perceived engagement and satisfaction. Importantly, comprehension was closely linked to sustained attention rather than interface type alone. These findings highlight the limits of interface-centered consent approaches and suggest that effective consent design must account for attention dynamics and selective engagement, rather than assuming that improved layout alone ensures understanding. 

\end{abstract}



\received{20 February 2007}
\received[revised]{12 March 2009}
\received[accepted]{5 June 2009}

\maketitle

\section{Introduction}

Digital platforms rely on privacy policies to legitimize the collection and use of personal data. These policies are intended to support informed consent, yet decades of research have shown that they are rarely read in full \cite{obar2018, ermakova2014}. Instead, they reduce consent to a rapid click rather than meaningful engagement.

While prior work has extensively documented \textit{why} users do not read privacy policies \cite{Sanchez2009, Leung2020, Harm2015}, less is known about \textit{how} this failure unfolds during interaction. Much of the existing literature relies on policy analysis, surveys, or interviews \cite{VANDERSCHYFF2024, Waldman2018, Javed2024}, which offer limited visibility into users’ moment-to-moment engagement \cite{smith2011}. The role of interface design in shaping users’ attention during consent interactions remains underexplored. 

This study uses eye-tracking as an empirical lens to examine how users allocate attention when interacting with privacy policies presented through common interface patterns (e.g., long-form scrolling vs. structured disclosure layouts). As a first step, we report findings from an exploratory study guided by the following research questions:

\begin{itemize}
\item \textbf{RQ1:} How do users allocate attention when interacting with different privacy policy interface patterns?
\item \textbf{RQ2:} What recurring patterns of engagement and disengagement emerge during these consent interactions?
\end{itemize}

By making visible the gap between governance assumptions about informed consent and users’ actual interaction behavior, this work contributes empirical evidence to ongoing discussions in HCI and privacy research about the limits of interface-mediated consent. We argue that while interface design can guide attention, it cannot by itself ensure meaningful understanding, especially in contexts involving complex data practices.

\section{Related Work}

\subsection{Privacy Policies}
Privacy policies are the primary means by which online platforms communicate data practices and obtain user consent. Rooted in Fair Information Practices (FIPs), notice-and-consent frameworks assume users can read disclosures, understand them, and make informed decisions \cite{Bennett1992, angst2009}. However, most users skip or skim policies and click “Agree” without reviewing them \cite{ermakova2014, FURNELL201212, Fabian2017}. Analyses of social media services show that terms average over 6000 words and require a high school junior reading level to comprehend \cite{Koebert2025}. 

Engagement is further constrained by structural power asymmetries. Access to digital services is typically contingent on acceptance, creating a forced-choice setting that normalizes habitual agreement \cite{KITKOWSKA2023}. Additionally, policies may change over time without salient notice \cite{SCHYFF2020}. Even when users attempt to engage, they lack the expertise to make informed judgments about data practice \cite{Aïmeur2013}. 

\subsection{Interface-mediated Engagement with Long-form Disclosures}
Navigation design patterns shapes how users access and process long-form text \cite{Cockburn2009, Harm2015}. Continuous scrolling supports rapid skimming \cite{Spool1997} but provides fewer stable anchors for locating content and can increase effort in text-heavy settings \cite{Bernard2002, Peytchev2006}. In contrast, structured disclosure layouts (e.g., collapsible sections) can improve overview and navigation efficiency \cite{shneiderman2010, KILSDONK2016}. It reduces on-screen text and minimizes cognitive load and completion time \cite{Zhang2005, Leung2020}. Providing short previews or summaries can further support selective reading by guiding attention and navigation decisions \cite{Cates2007, Burns2011, Aula2010}.

Eye-tracking has been widely used to study reading and attention during interaction \cite{Jacob1991, Rayner1998}. Fixation- and time-based measures can serve as behavioral indicators of attention allocation and processing effort during text engagement \cite{Iqbal2004, Nakayama2002, puma2018}. However, privacy-policy research has rarely examined how interface structure shapes these moment-to-moment behaviors. Although \citet{steinfeld2016} used eye-tracking to study whether default privacy policy presentation influences willingness to read, they did not compare how alternative interface structures shape processing effort. Our work extends this line by using eye-tracking to characterize attention allocation across common disclosure layouts, complemented by post-task questionnaire to contextualize users' interpretations.

\section{Method}

\subsection{Experimental Setup and Apparatus}

\begin{figure*}
    \centering
    \includegraphics[width=1\linewidth]{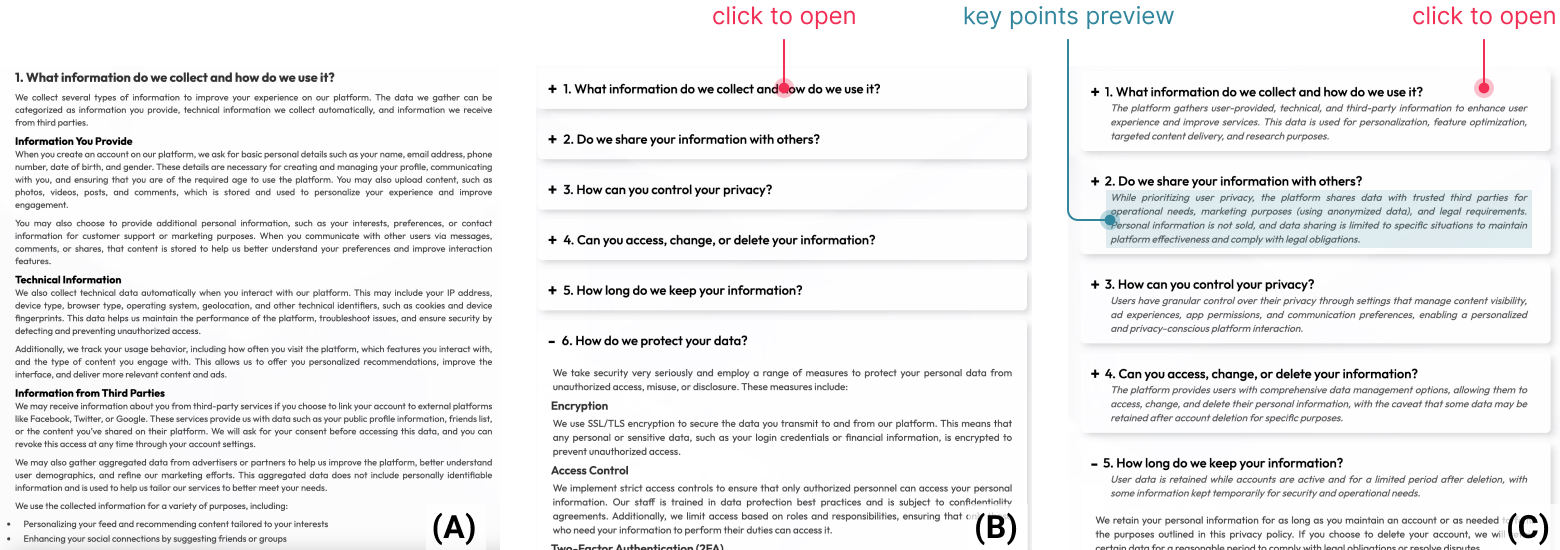}
    \caption{Privacy Policy Stimuli: (A) Plain (scrolling), (B) Tab (collapsible menus), (C) Highlight (collapsible menus with previews)}
    \label{fig:stimuli}
\end{figure*}

We created a standardized privacy policy and three interface variations grounded in established navigation and preview principles. The policy was synthesized from four social media platforms (TikTok, Instagram, Facebook, and Snapchat) to reflect common policy sections (e.g., data collection, sharing, and user rights). The final policy contained approximately 2,000 words. The text was implemented into three interactive webpage prototypes that differed only in navigation and preview features while keeping content, font, and visual style identical (see Figure~\ref{fig:stimuli}):

\begin{itemize}
    \item \textbf{Plain:} Continuous scrolling text with bold section headers.
    \item \textbf{Tab:} Content segmented into collapsible sections, displaying only section titles by default.
    \item \textbf{Highlight:} Collapsible sections with an additional one-sentence summary previewing the key content.
\end{itemize}

Eye movements were recorded using a screen-based Tobii X3-120 eye tracker integrated with iMotions (v10.1), sampling at 120 Hz. Participants viewed the stimuli on a 23-inch monitor (1920 $\times$ 1080 resolution). A standard 10-point calibration was conducted at the beginning of each session and repeated if necessary.

\subsection{Participants and Procedure}
52 participants were recruited through an online university research participation system (SONA) and received two course credits. Four cases were excluded due to technical issues or poor eye-tracking calibration, resulting in 48 valid datasets (age 18--43, $M$ = 20.64, $SD$ = 3.59; 41 women). Participants were randomly assigned to one of three interface conditions: Plain ($N$ = 17), Tab ($N$ = 15), and Highlight ($N$ = 16). 

The study protocol was reviewed for ethical compliance. Each session lasted approximately 30--45 minutes. Participants were told that the study examined the usability of a new social media interface. After completing a mock sign-up task, the assigned privacy policy interface appeared and eye movements were recorded. No entered information was stored. Participants then completed a brief filler task followed by a post-task questionnaire, and were fully debriefed.

\subsection{Data Processing and Analysis}

Eye-tracking data were exported using iMotions' open-source R-notebooks (iMotions A/S, 2025). Paragraph-level Areas of Interest (AOIs) were defined for the policy text. Key metrics included fixation- and time-based measures, such as fixation count, dwell time (ms), and first-fixation duration (FFD). Data with insufficient tracking quality (AOI ValidData < 70) were excluded before analysis.

The survey assessed participants' perceived engagement, comprehension, and attitudes. It included self-report items and a knowledge check (binary-coded: 1 = correct, 0 = incorrect) aligned with the policy content. Analyses focused on comparing attention patterns and self-reported outcomes across interface conditions.

\section{Results}

\subsection{Attention Allocation and Reading Behavior}

Eye-tracking data show that interface structure systematically shaped visual attention during policy reading. Participants in the Plain condition spent the longest time attending to the text (Mean dwell time = 7974 ms, $SD$ = 16295), followed by the Highlight condition ($M$ = 6213 ms, $SD$ = 8629), while the Tab condition elicited the shortest dwell time ($M$ = 5879 ms, $SD$ = 16293). The significant standard deviation suggests that some read the policy carefully, while others skimmed or quickly abandoned it. A Kruskal–Wallis test indicated a significant effect of interface condition on dwell time ($\chi^2$(2) = 15.49, $p < .001$). Post-hoc comparisons showed that the Highlight layout elicited significantly shorter dwell time than both Plain ($Z$ = 2.51, $p$ = .036) and Tab ($Z$ = 3.87, $p$ < .001). Plain and Tab did not differ significantly ($Z$ = 1.60, $p$ = .33).

Fixation-based measures revealed a similar pattern. Fixation count was strongly correlated with dwell time ($\rho = .98, p < .001$), indicating sustained visual engagement rather than sporadic glances. Mean fixation duration (MFD), an indicator of processing effort, differed across conditions ($F$(2,45) = 4.51, $p = .016$). Participants in the Tab condition exhibited significantly shorter fixations than those in Plain.

\begin{figure*}
  \centering
  \includegraphics[width=1 \linewidth]{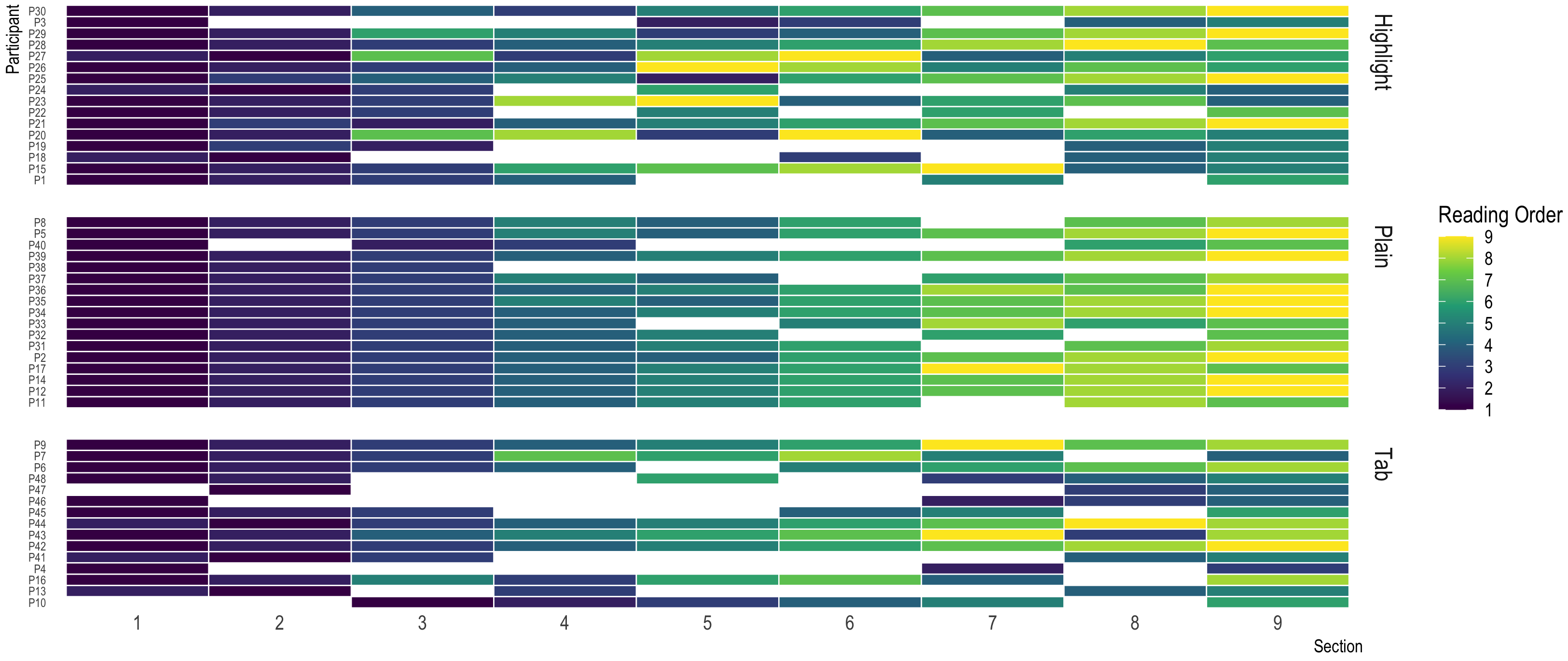}
  \caption{Reading Order Across Conditions. Rows represent participants and columns represent policy sections (AOI 1–9). Tile colors indicate the sequence in which each section was read (1 = first, 9 = last); gray tiles represent skipped sections.}
\label{fig:readingorder}
\end{figure*}

Reading-order visualizations (Figure~\ref{fig:readingorder}) further showed distinct navigation patterns across interfaces. In the Plain condition, participants generally followed a linear, top-to-bottom reading sequence, though attention frequently declined later. The Tab layout, by contrast, produced the most fragmented reading patterns, with frequent skipping and non-linear jumps between sections. The Highlight layout showed more coherent reading with fewer skipped sections.

\subsection{Cognitive Effort and Comprehension}

Interface design did not produce a clear group-level advantage in comprehension. Overall accuracy on the knowledge check averaged 63\% ($SD$ = 0.48), with item-level variation (e.g., 87.5\% for personalized advertising vs. 27.1\% for two-factor authentication). An ordinal logistic regression revealed no significant main effect of interface condition on comprehension ($\beta$ = 0.2109, $p > .05$).

At the individual level, comprehension was associated with visual engagement. Longer dwell time showed a moderate positive correlation with accuracy($r = .33, p < .001$), 
indicating that sustained engagement with policy content was associated with better recall. By contrast, mean fixation duration was weakly and negatively correlation with accuracy ($r = -.10, p = .043$), suggesting that prolonged individual fixations reflected processing difficulty.

\subsection{User Evaluations and Perceived Experience}

Post-task ratings revealed clear differences in how participants evaluated the interfaces. Survey results showed significant effects of interface condition on overall impression, satisfaction, stimulation, and perceived control (all $p < .05$; Table~\ref{tab:survey_privacy_policy}). Across these measures, the Tab layout consistently received the most positive evaluations, followed by Highlight, while the Plain layout was rated least favorably. 

Despite these differences, participants reported moderate levels of carefulness (all means $\approx$ 5 on a 10-point scale, $p > .05$). In other words, perceived attentiveness did not vary by layout, even though eye-tracking data showed substantial differences in actual attention allocation. This discrepancy points to a mismatch between perceived engagement and observed reading behavior.

\begin{table*}[t]
\centering
\caption{Descriptive and Inferential Statistics for User Evaluations Across Interface Conditions}
\label{tab:survey_privacy_policy}
\begin{tabular}{p{5.1cm}ccccccc}
\toprule
\textbf{Measure} & \multicolumn{2}{c}{\textbf{Plain}} & \multicolumn{2}{c}{\textbf{Tab}} & \multicolumn{2}{c}{\textbf{Highlight}} & \textbf{Test Statistic} \\ 
\cmidrule(lr){2-3} \cmidrule(lr){4-5} \cmidrule(lr){6-7}
 & M & SD & M & SD & M & SD & \\ 
\midrule
Terrible-Wonderful* & 5.53 & 1.07 & 6.59 & 1.50 & 5.73 & 1.17 & $F$(2, 45) = 3.81, $p$ = .03, $\eta^2$ = .14 \\
Frustrating-Satisfying* & 5.12 & 1.93 & 7.25 & 1.34 & 5.35 & 1.58 & $F$(2, 45) = 8.26, $p$ < .001, $\eta^2$ = .27 \\
Dull-Stimulating* & 4.12 & 1.93 & 6.31 & 1.58 & 3.88 & 1.69 & $F$(2, 45) = 11.1, $p$ < .001, $\eta^2$ = .33 \\
Sense of Control* & 6.11 & 1.63 & 6.82 & 1.37 & 5.07 & 1.56 & $F$(2, 45) = 5.09, $p$ = .01, $\eta^2$ = .18 \\
\bottomrule
\end{tabular}
\begin{tablenotes}
\footnotesize
\item \textit{Note.} *Significant differences between interface conditions ($p$ < .05).
\end{tablenotes}
\end{table*}

\section{Discussion}

The results reveal a consistent pattern across behavioral and self-report measures: interface structure shaped how users allocated attention and experienced privacy policies, but these differences did not translate into uniform gains in comprehension. In particular, the Highlight interface supported more efficient and coherent reading patterns, while the Tab layout elicited the highest in-task satisfaction and stimulation.

This pattern can be interpreted through processing fluency. Prior work suggests that information that is easier to process may feel less stimulating in the moment, even while producing greater comfort and positive evaluation overall \cite{Reber2004, Alter2009}. In this study, concise previews in the Highlight appeared to reduce cognitive friction and guide attention, enabling users to navigate complex policy content more efficiently. Meanwhile, this fluency may have reduced the perceived effort associated with reading, contributing to lower ratings of stimulation and satisfaction compared to the more interactive Tab layout. By contrast, Tab afforded users a stronger sense of agency through clicking and exploration, heightening perceived control and engagement even when reading behavior was more fragmented \cite{Liu01122002, Sundar2008}.

The discrepancy between self-reported carefulness and observed behavior further underscores the limits of relying on subjective evaluations in consent contexts. Participants across conditions reported similar levels of attentiveness, yet eye-tracking data revealed substantial variation in attention allocation and reading behavior. This gap suggests that users may overestimate their engagement with privacy policies, echoing prior findings that individuals often misjudge their privacy knowledge and level of informed consent \cite{Ma2023, wang2025}. Interfaces that feel engaging or manageable may therefore reinforce a false sense of understanding without ensuring sustained attention to policy content.

Importantly, comprehension was not driven by interface type alone. Instead, understanding was closely associated with how attention was distributed during reading. Longer overall engagement was linked to higher comprehension, whereas prolonged individual fixations appeared to reflect processing difficulty rather than deeper understanding. In this sense, the Highlight layout functioned less as a direct cognitive aid and more as an attentional scaffold: it shaped how reading effort was allocated, allowing attentions to translate more efficiently into understanding.

These findings help explain why prior attempts to redesign privacy policies, such as nutrition labels \cite{Kelley2009, Kelley2010}, comics \cite{Tabassum2018}, and condensed summaries \cite{Bahrini2022}, have improved perceived usability without producing consistent comprehension gains \cite{Reinhardt2021}. While such interventions enhance visual accessibility, they cannot generate motivation where it is absent. Interface design alone cannot guarantee informed consent; rather, it influences \textit{how} users engage once they choose to attend. From this perspective, effective consent design requires balancing visual comfort, perceived agency, and attentional guidance, while recognizing the persistent motivational barriers that shape real-world privacy behavior.

\section{Design Implications}

Our findings suggest consent interfaces should support goal-oriented engagement while reducing cognitive friction without fostering a premature sense of understanding.Users rarely seek exhaustive understanding; instead, they look for answers to specific concerns. Structured navigation, previews, and layered disclosure can help users locate relevant information efficiently. Additionally, positive user experience alone should not be equated with effective consent. Designing for informed consent therefore requires balancing comfort, perceived agency, and attentional guidance, acknowledging that informed consent is not a single interaction but an ongoing, context-dependent process.

\section{Limitations and Future Work}

This study has several limitations. First, it examines only two design features, focusing on how structural disclosure mechanisms shape attention during consent interactions. Future work could extend this approach to other elements, such as visual emphasis or color, and examine how they interact with navigational structure. Second, the participant sample consisted mainly of university students, which may limit the generalizability of the results. Although this group represents frequent social media users, individuals with different levels of digital literacy or cultural backgrounds may engage with privacy policies in distinct ways. Finally, while eye-tracking measures offer valuable insights into attention allocation, they provide indirect indicators of cognitive processing. Future work could integrate physiological or longitudinal approaches to capture deeper cognitive and behavioral dynamics.

\section{Conclusion}

This study shows that engagement with privacy policies emerges from the interaction between interface, attentional effort, and motivation. Interface navigation alone does not ensure comprehension. Instead, understanding depends on whether interface cues are able to sustain attention and motivate users choose to engage.

These findings highlight the limits of interface-centered approaches to consent and suggest that effective consent design requires balancing visual comfort, perceived agency, and attentional guidance. This work underscores the need to design consent interfaces that acknowledge selective engagement and the persistent motivational challenges surrounding privacy policy reading. This study contributes empirical evidence to ongoing discussions in HCI and privacy research about how interface design shapes, but does not guarantee, informed consent.

\begin{acks}
We would like to thank Professor Kevin Wise for his guidance and for providing access to the eye-tracking lab and equipment that made this research possible.

\end{acks}

\bibliographystyle{ACM-Reference-Format}
\bibliography{sample-base}

\end{document}

%% file: questionsformatting.tex
\newlist{compactitem}{itemize}{5}
\setlist[compactitem]{leftmargin=*, nosep}
\setlist[compactitem, 1]{label=\textbullet}
\setlist[compactitem, 2]{label=\textendash}
\setlist[compactitem, 3]{label=\textasteriskcentered}
\setlist[compactitem, 4]{label=\textperiodcentered}

\makeatletter
\let\orgItem\item
\NewDocumentCommand\fixedItem{ o }{%
   \IfNoValueTF{#1}%
      {\orgItem}
      {\orgItem[#1]\def\@currentlabel{#1}}
}
\makeatother

\newlist{questions}{enumerate}{1}
\setlist[questions]{align=left, labelwidth=1.5em, labelsep=.5em, listparindent=0pt, itemindent=0.5em, leftmargin=0.8em, before=\let\item\fixedItem, label=Q\arabic*}


\setlength\multicolsep{0pt}
\newlist{answers}{itemize}{1}
\setlist[answers]{leftmargin=*, nosep, align=left, label=$\bigcirc$}
\newlist{answers*}{itemize*}{1}
\setlist[answers*]{label=$\bigcirc$}

\newcommand{\head}[1]{{{\textbf{#1}}}} 
\newcommand{\paraSurvey}[1]{{\small{\par{#1}}}} 
\newcommand{\condition}[1]{{\small{\par{\color{orange} If #1}}}}
\newcommand{\fillin}[1]{{\fbox{#1}}}
\newcommand{\fillinessay}{\fbox{\parbox{.75\linewidth}{~}}}

\newlist{scale}{enumerate}{1}
\setlist[scale]{
  label=\textbf{\arabic*},  
  labelsep=0.5em,
  itemsep=0pt,
  leftmargin=2.5em,
  labelwidth=1.5em,
  align=left,
  itemindent=0pt,
  listparindent=0pt
}

\newcommand{\scalequestion}[2]{
  \paraSurvey{\textbf{#1}} \\[0.5em]
  \begin{scale}
    \setcounter{enumi}{#2} 
    \item[0] \hspace{0.5em}%
      \foreach \x in {0,...,9}{\makebox[1em][c]{\x}} 
  \end{scale}
}

\newcommand{\scalerow}[1]{
  \begin{center}
    \foreach \x in {0,...,9}{\textbf{\x}\hspace{1em}}
  \end{center}
}

\newlist{subquestions}{enumerate}{1}
\setlist[subquestions]{
  label*=\alph*),
  leftmargin=2em,
  itemsep=3pt,
  topsep=3pt,
  before*=\smallskip,
}